\documentclass[amsmath,amssymb,12pt]{article}
\usepackage[all]{xy}
\usepackage[centertags]{amsmath}
\usepackage{latexsym}
\usepackage{amsfonts}
\usepackage{amssymb}
\usepackage{amsthm}
\usepackage{graphicx}
\def\R{{\rm I\!R}}
\newcommand{\beq}{\begin{eqnarray}}
\newcommand{\eeq}{\end{eqnarray}}

\begin{document}


\title{Tessellated and stellated invisibility}

\author{Andr\'e Diatta,$^{1}$ Andr\'e Nicolet,$^{2}$ S\'ebastien Guenneau,$^{1}$
and Fr\'ed\'eric Zolla$^{2}$}
\maketitle
\date{}
\noindent
{\footnotesize
\thanks{$^{1}$Department of Mathematical Sciences, Peach Street, Liverpool L69 3BX, UK\\ Email addresses: adiatta@liv.ac.uk; guenneau@liv.ac.uk
\\
$^{2}${\em{Institut Fresnel, UMR CNRS
6133, University of Aix-Marseille III,
\newline\noindent
case 162, F13397 Marseille Cedex 20, France\\
Email addresses: andre.nicolet@fresnel.fr; frederic.zolla@fresnel.fr}};
\\

}
}

\begin{abstract}
We derive the expression for the anisotropic heterogeneous matrices
of permittivity and permeability associated with two-dimensional
polygonal and star shaped cloaks. We numerically show using finite
elements that the forward scattering worsens when we increase the
number of sides in the latter cloaks, whereas it improves for the former ones.
This antagonistic behavior is discussed using a rigorous asymptotic approach.
We use a symmetry group theoretical approach to derive the cloaks design.
\end{abstract}
\noindent
{\bf \underline{Ocis}:} {\em (000.3860) Mathematical methods in physics; (260.2110)
Electromagnetic theory; (160.3918) Metamaterials; (160.1190)
Anisotropic optical materials}


\section{Introduction}
Back in 2006, a team led by Pendry and Smith theorized and subsequently experimentally
validated that a finite
size object surrounded by a spherical coating consisting of a metamaterial
might become invisible for electromagnetic waves \cite{pendrycloak,cloakex}.
Many authors have since then dedicated a fast growing amount of work to the invisibility cloaking problem. However, there are alternative approaches,
 including inverse problems's methods \cite{greenleaf} or plasmonic properties of negatively refracting index materials \cite{milton2}.
It is also possible to cloak arbitrarily shaped regions, and for cylindrical domains of arbitrary cross-section, one can generalize \cite{pendrycloak}
 to linear radial geometric transformations \cite{nzg2008}
\begin{equation}
\left\{
\begin{array}{ll}
r' &= R_1(\theta)+r(R_2(\theta)-R_1(\theta))/R_2(\theta) \; , \; 0\leq r\leq R_2(\theta) \; ,\\
\theta' &= \theta \; , \; 0<\theta\leq 2\pi \; ,\\
x_3' &= x_3 \; , \; x_3 \in \R \; ,
\end{array}
\right.
\label{transfo}
\end{equation}
where $r'$, $\theta'$ and $x_3'$ are ``radially contracted
cylindrical coordinates'' and $(x_1,x_2,x_3)$ is the Cartesian
basis. This transformation maps the arbitrary domain $D_{R_2(r,\theta)}$ onto the coating $D_{R_2(r,\theta)}\setminus D_{R_1(r,\theta)}$ with $R_1(r,\theta)< R_2(r,\theta)$.
 In other words, if a source located in $\R^2 \setminus
D_{R_2(r,\theta)}$ radiates in a vacuum, the electromagnetic field
cannot reach the domain $D_{R_1(r,\theta)}$ and therefore this
region is a shelter for any object. Moreover, the transformation
maps the field within the domain $D_{R_2(r,\theta)}$ onto itself by
the identity transformation.
\begin{figure}[h!]
\centerline{
\includegraphics[width=15cm,angle=0]{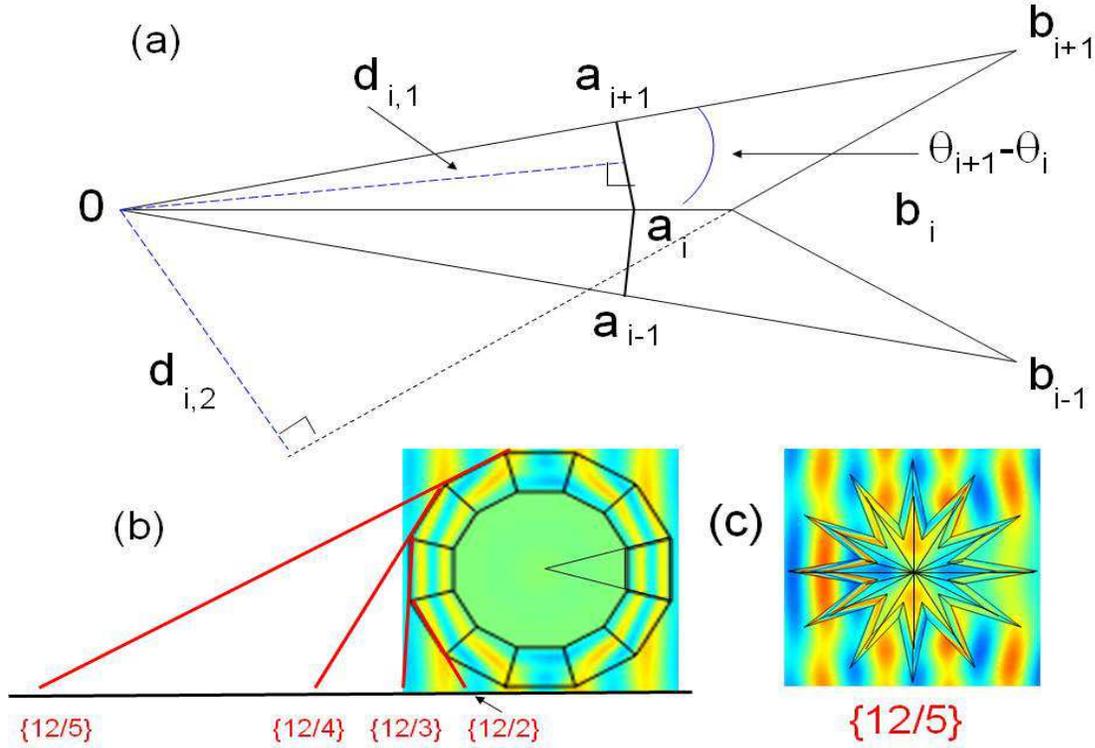}}
\caption{(a) Triangle of vertices $(0,0)$,
$(b_i,\theta_i)$ and $(b_{i+1},\theta_{i+1})$ used in the
triangulation of a polygon (see (b)) or a polygram (see (c)) mapped
under Eq. (\ref{transfo}) onto the trapezoid of vertices
$(a_i,\theta_i)$, $(b_{i},\theta_{i})$, $(b_{i+1},\theta_{i+1})$ and
$(a_{i+1},\theta_{i+1})$. The angles $\theta_i$ and the segments
$d_{i,j}$ are as in Eq. (\ref{dedetrick}); (b) Dodecagon and first step
in its four tessellations $\{12/2\}$-$\{12/5\}$; (c) Tessellation
$\{12/5\}$ and representative samples of its triangulation.}
\label{figgeo}
\end{figure}
In this paper, we would like to investigate designs of polygonal and star-shaped cloaks which we can triangulate, see Figs. \ref{figgeo},  \ref{fig2},  \ref{fig3}.
In this case, consider a given triangle, numbered $i$, of the triangulation of the region we want to cloak. We adapt Eqs. (\ref{transfo}) to
 the specific geometry of triangles, to derive a geometric transform  that maps the whole triangle of vertices $(0,0)$, $(b_{i},\theta_i)$,
 $(b_{i+1},\theta_{i+1})$ onto  a trapezoidal region of vertices $(a_{i},\theta_i)$, $(a_{i+1},\theta_{i+1})$, $(b_{i},\theta_i)$, $(b_{i+1},\theta_{i+1})$, see
Fig. \ref{figgeo}.
This means that in Eqs. (\ref{transfo}), the points $(R_1(\theta),\theta)$ and $(R_2(\theta),\theta)$  which now respectively lie in the line segments
 $ [(a_{i},\theta_i),(a_{i+1},\theta_{i+1})]$ and $[(b_{i},\theta_i),(b_{i+1},\theta_{i+1})],$  are referred to as
$(R_{i,1}(\theta),\theta)$ and $(R_{i,2}(\theta),\theta)$ where
\begin{equation}
\left\{
\begin{array}{lr}
R_{i,j} = d_{i,j}/\cos(\theta-\theta_0), ~~ \theta_i\leq\theta\leq \theta_{i+1}, ~~j=1,2 \nonumber \\
d_{i,j} = c_{i,j} c_{i+1,j} \sin(\theta_{i+1}-\theta_i)/ \sqrt{c_{i,j}^2+c_{i+1,j}^2- 2 c_{i,j}c_{i+1,j} \cos(\theta_{i+1}-\theta_i)} , ~~ j=1,2 \nonumber \\
\theta_0 =  \theta_i+\arccos({d_{i,j}}/{c_{i,j}})=\theta_{i+1}+\arccos({d_{i,j}}/{c_{i+1,j}}) \nonumber\\
c_{i,1} = a_{i} \; , \; c_{i,2} =b_{i}
\end{array}
\right.
\label{dedetrick}
\end{equation}
In the sequel, we make use of symmetries to further simplify the implementation of such cloaks.
More precisely, we work in a reference triangle and construct the overall cloak via compound reflection symmetries and rotations from this reference region.

\section{Change of coordinates and pullbacks from optical to physical space}
The basic principle of our approach is to transform a geometrical
domain into another one and to search how the Maxwell equations to
be solved have to be changed. As we start with a given set of
equations on a given domain, it seems at first sight that we have to
map this domain on a new one. Nevertheless, it is the opposite that has to be done. On the original domain, say
$\Omega$, the equations are described in a particular coordinate
system $x_i$ (taken here by default to be Cartesian
coordinates). We want to establish a one-to-one correspondence
with a new domain $\Omega'$ where we use a general (possibly
non-orthogonal) coordinate system $y_i .$
\begin{equation}
\begin{array}{lcl}
\mathrm{domain\, \Omega\; with} & \longleftarrow & \mathrm{domain\, \Omega'\; with}  \\
\mathrm{coordinates}\; x_i & x_i(y_j) & \mathrm{coordinates}\; y_i   \\
\\
\mathrm{differential}\; \mathrm{d}x_i  & \longrightarrow &
\mathrm{differential}\; \sum_j \frac{\partial x_i(y_j)}{\partial
y_j}\mathrm{d}y_j
\\
& \mathrm{pullback}\\
\end{array}
 \label{pullback}
\end{equation}
More precisely, consider a map from a coordinate system
$\{y_1,y_2,y_3\}$ to the initial one $\{x_1,x_2,x_3\}$ given by the
transformation $x_1(y_1,y_2,y_3)$, $x_2(y_1,y_2,y_3)$ and
$x_3(y_1,y_2,y_3)$. Here, we must point out that, we are mapping the
transformed domain and coordinate system onto the initial one with
Cartesian coordinates. This change of coordinates is characterized
by the transformation of the differentials through $\mathrm{d}{\bf
x}=\mathbf{J}_{xy}\mathrm{d}{\bf y}$ where
\beq
 \mathbf{J}_{xy}=
\frac{\partial(x_1,x_2,x_3)}{\partial(y_1,y_2,y_3)},
~\mathrm{d}{\bf x}=(\mathrm{d}x_1,\mathrm{d}x_2,\mathrm{d}x_3)^T
\text{~and~} \mathrm{d}{\bf
y}=(\mathrm{d}y_1,\mathrm{d}y_2,\mathrm{d}y_3)^T.\eeq
 In the sequel we
will also need a compound transformation. Let us consider three
coordinate systems $\{y_1,y_2,y_3\}$, $\{z_1,z_2,z_3\}$, and
$\{x_1,x_2,x_3\}$ (possibly on different regions of spaces). The two
successive changes of coordinates are given by the Jacobian matrices
$\mathbf{J}_{xz}$ and $\mathbf{J}_{zy}$. This rule naturally applies
for an arbitrary number of coordinate systems and we call the
resulting Jacobian matrix
$\mathbf{J}=\mathbf{J}_{xz}\mathbf{J}_{zy}.$ In electromagnetism,
changes of coordinates amount to replacing the different materials
(often homogeneous and isotropic, which corresponds to the case of
scalar piecewise constant permittivity $\varepsilon$ and
permeability $\mu$) by equivalent inhomogeneous anisotropic
materials described by a transformation matrix ${\bf T}$ (metric
tensor) \cite{ward,nicolet,zgnp07,Ulf} so that the new
permittivities and permeabilities  read
\begin{equation}
\underline{\underline{\varepsilon'}} =\varepsilon \mathbf{T}^{-1} \;
,  \quad
 \hbox{and} \quad
\underline{\underline{\mu'}}=\mu \mathbf{T}^{-1} \hbox{ where
$\mathbf{T} \!= \! \mathbf{J}^T
\mathbf{J}/\mathrm{det}(\mathbf{J})$} \; . \label{epsmuT}
\end{equation}

\section{Transformation optics and the geometry of irregular polygons}
\subsection{Material properties of the regions}
From Eqs. (\ref{transfo}), (\ref{dedetrick}) and (\ref{epsmuT}), we
derive the transformation matrix associated with a reference
trapezoidal region (see Fig. \ref{figgeo}) used to generate, via
reflections symmetry and rotations, two-dimensional polygonal and star-shaped
cloaks:
\begin{equation}
\mathbf{T}^{-1} =
\left(
\begin{array}{ccc}
 \frac{e_{12}^2+f_r^2}{e_{11} f_r r'} &- \frac{e_{12} }{f_r}   &0 \\
 - \frac{e_{12} }{f_r}& \frac{e_{11}r' }{f_r}   &0 \\
 0 &  0 &\frac{e_{11}f_r }{r'} \\
\end{array}
\right) \;
, \label{invt}
\end{equation}
where \beq
e_{11}(\theta') = R_2(\theta')/(R_2(\theta')-R_1(\theta'))\eeq
and
\beq e_{12}(r',\theta')=\displaystyle{\frac{(r'-R_2(\theta'))R_2(\theta')\frac{dR_1(\theta')}{d\theta'}
-(r'-R_1(\theta'))R_1(\theta')\frac{dR_2(\theta')}{d\theta'}}{(R_2(\theta')-R_1(\theta'))^2}}
\eeq
 for \beq
 R_1(\theta')\leq r'\leq R_2(\theta'),\eeq
  with
\beq
f_r(r',\theta')=(r'-R_1(\theta'))\frac{R_2(\theta')}{R_2(\theta')-R_1(\theta')}
\;
 \eeq
 (the expression for $e_{12}$ corrects that in \cite{nzg2008}). Elsewhere, $\mathbf{T}^{-1}$ reduces
to the identity matrix ($e_{11} =1$, $e_{12} =0$ and $f_r=r'$ for
$r'>R_2(\theta')$).
However, Eqs. (\ref{epsmuT}) and (\ref{invt}) can just be applied to derive the (anisotropic heterogeneous)
permittivity $\underline{\underline{\varepsilon'}}$ and permeability
$\underline{\underline{\mu'}}$ in the reference triangle. An elegant way to deduce optical properties
of other regions of the triangulation is to use linear groups of symmetry in the plane, as now explained.

\subsection{Symmetries and transformation groups in cloaks' construction }
Here, we use Euclidean groups, namely reflections symmetry and rotations in
the plane, as a tool to design very general polygonal and
star-shaped cloaks. We choose our reference triangle (labeled as
triangle number $1$, or just referred to as first triangle) to be in
the first quadrant with one of its sides belonging to the
$x_1$-axis. In the regular polygons case, the triangle number $i$ of
the triangulation, is simply obtained from this first one, via the
rotation with angle $(i-1)\frac{\pi}{n}.$ In the irregular case
(star-shaped), due to geometric chirality properties (irregular
triangles are not identical to their mirror images), triangles of
odd numbers cannot be obtained by rotations from the original one.
More precisely, the neighboring triangles are obtained, from the
first one, as
 mirror images with respect to the shared side (more precisely, a mirror image
 with respect to the plane perpendicular to the triangle and containing the shared side).
 Hence, any triangle of even number $i,$ is obtained by applying to the first triangle
 the linear map
 \beq
 (x_1,x_2)\mapsto (x_1\cos (\frac{\pi}{n}i)+x_2\sin(\frac{\pi}{n}i), x_1\sin (\frac{\pi}{n}i)-x_2\cos(\frac{\pi}{n}i))
 \eeq
 which is the composition of the reflection
$(x_1,x_2)\mapsto (x_1,- x_2)$ with respect to $x_1$-axis, followed by the rotation with angle $\frac{\pi}{n}i.$ Whereas,
  every triangle of odd number $i,$ is obtained by rotating the first triangle with angle $(i-1)\frac{\pi}{n}.$
In short, optical properties of all regions
 within the triangulation are deduced from the reference trapezoidal region using
 compositions of rotations and reflections. This algorithm is particularly well-suited for all  polygonal or star shaped regions displaying $n$-fold symmetries.

\subsection{Tesselation and stellation}
We are interested in the antagonistic behaviour of polygonal and
star shaped cloaks. There is a natural correspondence between these
two types of cloaks, and this can be seen as follows.

Consider a
regular star polygon which is represented by its Schl\"afli symbol
$\{n/m\}$, where $n$ is the number of vertices, and $m$ is the step
used in sequencing the edges around it.
For example, take a dodecagram $\{12/5\}$ as in Fig. \ref{figgeo}
which has twelve vertices, see Fig. \ref{fig3}(c). One can draw a
straight line joining each of its vertices to another one, which is
five vertices apart. Or equivalently, choose a vertex, say, on the
horizontal axis and project the other vertices onto that axis as
depicted in Fig. \ref{figgeo}(b). For a given dodecagon with
Schl\"afli symbol $\{12\}$, there are four segments obtained by
projection which generate four distinct stellations
$\{12/2\}$-$\{12/5\}$. Note that a regular polygon with n sides has
$(n-4)/2$ stellations if $n$ is even, and $(n-3)/2$ stellations, if
n is odd. We display in Fig. \ref{fig2} and \ref{fig3} a gallery
of cloaks together with their Schl\"afli symbol.
We note that the higher the Schl\"afli symbol of the polygons, the
smoother the wave patterns (the less scattering).

On the contrary,
the scattering worsens for their corresponding stellations. This is
not surprising since it is fairly intuitive that a cloak with many
sharp angles will tend to display more singularities in the entries
of the transformation matrix describing the electromagnetic material
properties. However, we now give an asymptotic argument supporting
this numerical observation.
\subsection{Asymptotic behaviour of the $T^{-1}$ matrix}
 The tessellation of regular polygons
involves $n$ regular triangles. We can chose
\beq
a_i=a_{i+1}=:R_1, ~ b_i=b_{i+1}=:R_2, ~
 \theta_{i+1}-\theta_{i}=\frac{\pi}{n} ~\text{~and~} \theta=s\frac{\pi}{n}, ~0\leq s\leq 1.\eeq
  The coefficients of  $\mathbf{T}^{-1}$ now read \beq
  T^{-1}_{11}&=&\frac{r'-R_1}{r'}+O(\frac{1}{n^2}),\nonumber\\
   T^{-1}_{22}&=& \frac{r'}{r'-R_1}+O(\frac{1}{n^2}),\nonumber\\
 T^{-1}_{33}&=&(\frac{R_2}{R_2-R_1})^2\frac{r'-R_1}{r'}+O(\frac{1}{n^2}),\nonumber\\
 T^{-1}_{12}&=& T^{-1}_{21}=-\frac{\pi R_1(r'+2s(r'-R_1))}{2r'(r'-R_1)}\frac{1}{n}+O(\frac{1}{n^3}),\eeq
  and
$T^{-1}_{ij}=0,$ otherwise. So as $n$ grows, $\mathbf{T}^{-1}$ tends
to
\begin{equation}
\mathbf{T}_{\infty}^{-1} = \hbox{Diag}(\frac{r'-R_1}{r'},\frac{r'}{r'-R_1},
{(\frac{R_2}{R_2-R_1})}^2\frac{r'-R_1}{r'}).
\label{invtinfty}
\end{equation}
This is the expression of the transformation matrix for a circular
cloak, see also \cite{delustrac} for a particular case. Now assume
 $a_{i+1}\neq a_i,$ or $b_{i+1}\neq b_i$, which is expected to hold
in star shaped case, the Taylor expansion shows that, at least one $T^{-1}_{ij}$ tends to infinity or
det$(\mathbf{T}^{-1})$ tends to zero, as $n$ increases.
 For example,
  \beq
 T^{-1}_{11}=\frac{-(a_{i+1}-a_i)^2a_i^2a_{i+1}^2}
{(r's(a_{i+1}-a_i)+a_{i+1}(r'-a_i))(a_{i+1}
 +s(a_{i+1}-a_i))^3\pi^2r'}n^2+O(1),\eeq
  if $a_{i+1}>a_i$ and $b_{i+1}> b_i.$
This explains the origin of the singularities in the material parameters
of stellated cloaks.
\section{Finite elements computations}
\begin{figure}[h!]\centerline{
\includegraphics[width=15cm]{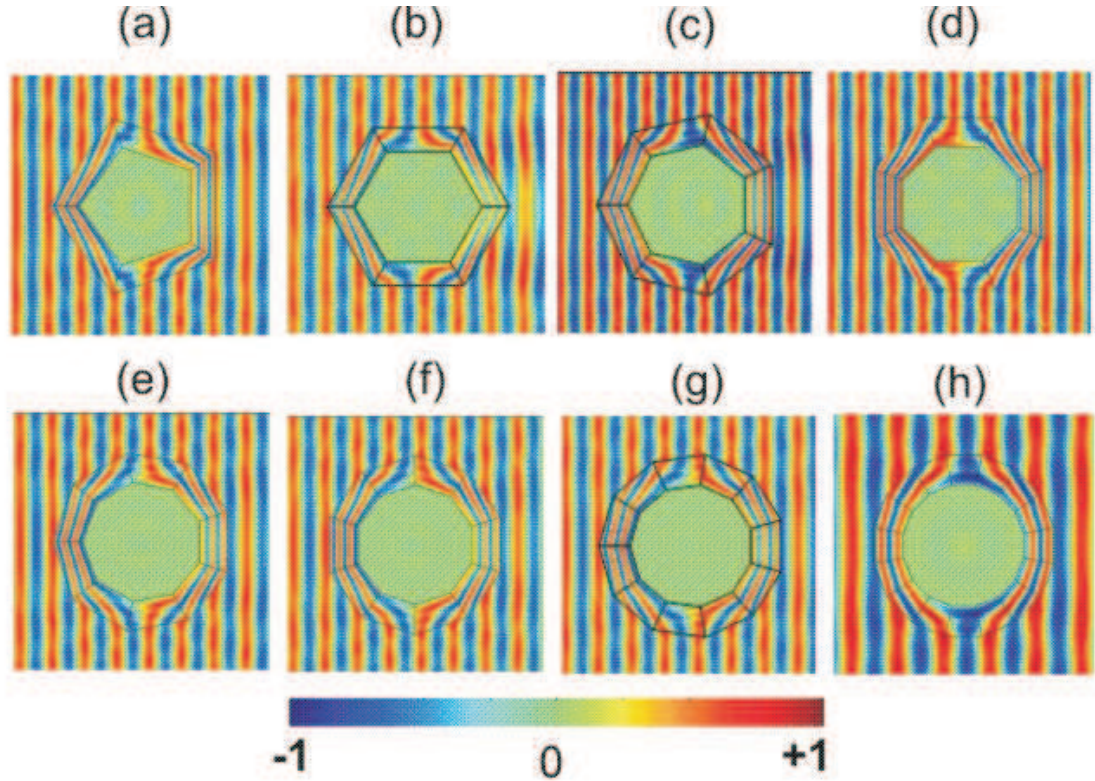}}
\caption{Real part of the longitudinal component  $E_3$ of the
electric field scattered by a (a) pentagon $\{5\}$, (b) hexagon
$\{6\}$, (c)  heptagon $\{7\}$, (d) octagon $\{8\}$, (e)
nonagon $\{9\}$, (f) decagon $\{10\}$, (g) hendecagon $\{11\}$
(h) hexadecagon $\{16\}$ polygonal cloaks. Both forward and
backward scattering for a plane wave propagating from the left are
nearly vanishing.} \label{fig2}
\end{figure}

\begin{figure}[h!]\centerline{
\includegraphics[width=14cm]{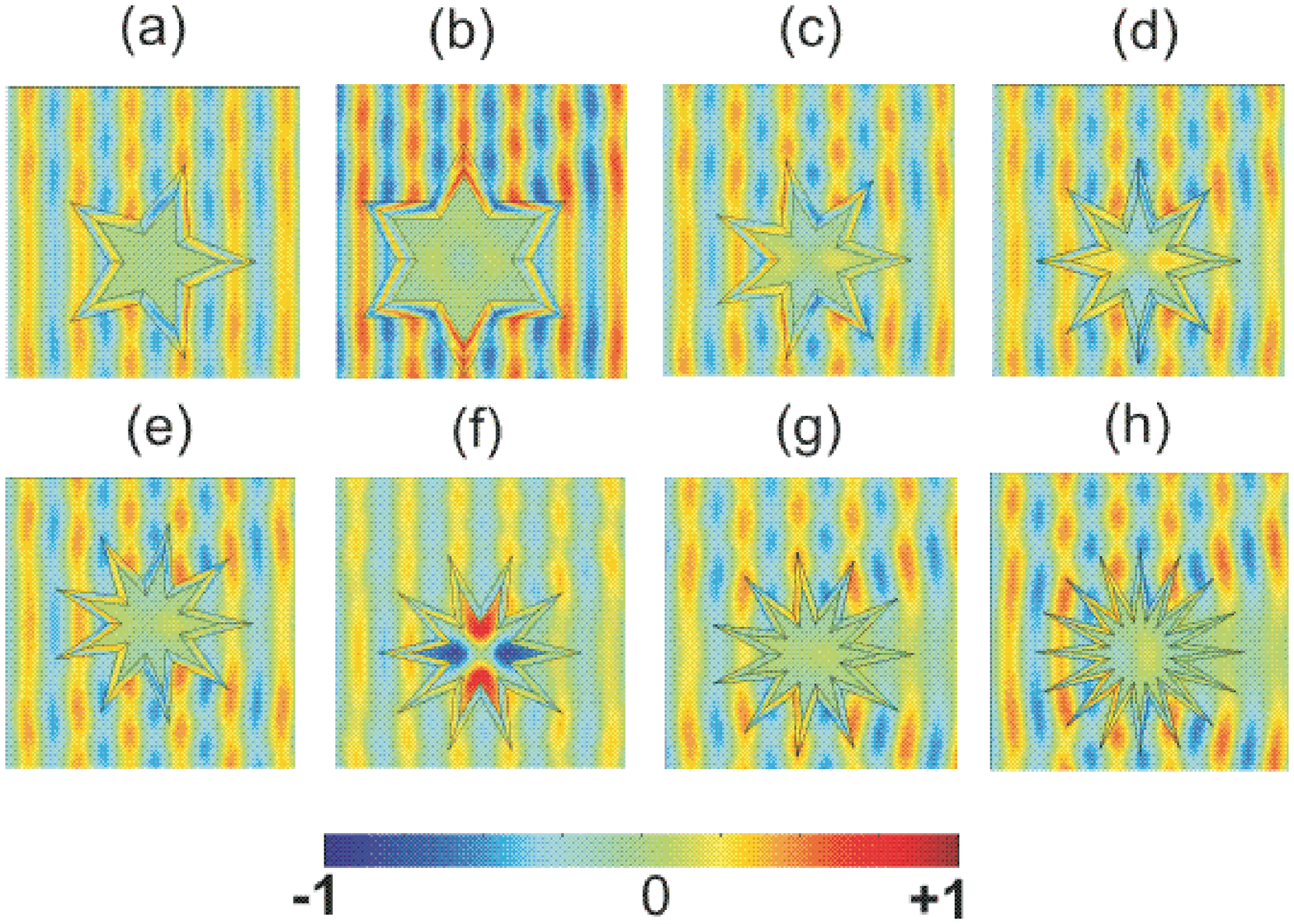}}
\caption{Real part of the longitudinal component  $E_3$ of the
electric field scattered by a (a) pentagram $\{5,2\}$, (b)
hexagram $\{6,2\}$, (c) heptagram  $\{7/3\},$ (d) octagram  $
\{8/3\}$, (e) nonagram $ \{9/4\}$, (f) decagram $ \{10/4\}$,
(g) hendecagram $ \{11/5\}$ (h) hexadecagram $ \{16/7\}$
star-shaped cloaks. Unlike for computations reported in Figure 2,
 scattering worsens from panels (a) to (h), and most specifically forward scattering.}
\label{fig3}
\end{figure}

Let us now compute the total electromagnetic field for a plane wave
incident upon a cylindrical cloak. In $p$ polarization, if ${\bf
e}_3$ is a unit vector oriented along the axis of the cylinder, the
longitudinal electric field ${\bf E}_l=E_3(x_1,x_2) {\bf e}_3$ is
solution of:
\begin{equation}\label{eq:El}
\nabla\times\left( \underline{\underline{\mu'}}^{-1}\nabla
\times{\bf E}_l \right) - k^2\underline{\underline{\varepsilon'}}
{\bf E}_l={\bf 0}
\end{equation}
where $k=\omega\sqrt{\mu_0\varepsilon_0}=\omega/c$ is the wavenumber, $c$ being the speed of light in
vacuum, and $\underline{\underline{\varepsilon'}}$ and
$\underline{\underline{\mu'}}$ are defined by Eqs. (\ref{epsmuT}) and
(\ref{invt}). Also, ${\bf E}_l={\bf E}_i+{\bf E}_d$, where ${\bf E}_i=e^{ikx_1}{\bf e}_3$ is the incident field and ${\bf E}_d$ is the diffracted
 field which satisfies the usual outgoing wave condition
(to ensure existence and uniqueness of the solution). We have
implemented the weak form of this scattering problem in the finite
element package COMSOL (with Perfectly Matched Layers modelling the
unbounded domain). We consider a plane wave incident from the left
at wavelength $\lambda=0.35$ (\emph{all lengths are given in
arbitrary units, $\mu m$ for instance}). We note that when we
increase the number of sides of a polygonal cloak, the scattering
becomes less apparent in Fig. \ref{fig2}, which is in agreement with
the fact that the transformation matrix ${\bf T}^{-1}$ tends to that
of a circular cloak, see formula (\ref{invtinfty}). On the contrary,
both forward and backward scattering worsen when we increase the
number of sides of corresponding stellated (star-shaped) cloaks, see
Fig. \ref{fig3}, and we have identified the origin of this singular
behaviour in the previous section. Note that we kept the frequency
constant as well as the mesh (with around $60.000$ elements), so as
to exemplify that any attempt to manufacture a star-shaped cloak
with many sides would be fairly challenging: when we refine the
mesh, the scattering indeed reduces suggesting severer constraints
on the corresponding metamaterial. Interestingly, the field can be
confined when the frequency of the plane wave excites some resonance
of the star-shaped cavity inside the cloak, see Fig. \ref{fig3} (f).
\section{Conclusion}
Analysis of polygonal invisibility cloaks is of current interest
\cite{delustrac,zhang,chao}. In this paper, we made use of a symmetry group action
 to construct such cloaks, and we have numerically shown
that, for a given frequency, $n$-sided polygonal cloaks behave in a
way similar to circular cloaks when $n$ increases whereas their
stellated counterparts become more and more singular. An asymptotic
algorithm supports these observations. A natural question to ask is
whether it is possible to design such cloaks using a homogenization
approach. Our analysis suggests that the answer should be positive
for the former type of cloaks (as their boundary smoothens for large
$n$ and as it has been shown that structured circular cloaks can be
homogenized \cite{farhat}). We conjecture that the answer should be
negative for the latter cloaks: it is well-known that homogenization
theory fails in so-called bad domains that display boundaries with
sharp corners (for which star-shaped invisibility cloaks is an
electromagnetic paradigm).

\section*{Acknowledgments}
A. Diatta and S. Guenneau acknowledge funding from EPSRC grant
EP/F027125/1.

\end{document}